\documentclass[prx,aps,letterpaper,twocolumn,allowfontchageintitle=true,unpublished]{quantumarticle}

\usepackage{bbm}
\usepackage{dcolumn}
\usepackage{bm}
\usepackage{graphicx}
\usepackage{hyperref}
\usepackage{color}
\usepackage{version}
\usepackage{mathtools}
\graphicspath{{figures/}}
\usepackage{natbib}
\usepackage{multicol}

\usepackage{float}
\usepackage{wasysym}
\usepackage{tabularx}

\usepackage[table]{xcolor}
\usepackage{booktabs}
\usepackage{xcolor}
\usepackage{braket}

\begin{document}

\title{\centering \color{quantumviolet} Quantum schoolbook multiplication with fewer Toffoli gates  \newpage}
\author{\vspace{-7ex}}
\affiliation{Daniel Litinski @ PsiQuantum, Palo Alto}
\date{\vspace{-7ex}}
{\centering \maketitle}

\begin{abstract}

This paper presents a method for constructing quantum circuits for schoolbook multiplication using controlled add-subtract circuits, asymptotically halving the Toffoli count compared to traditional controlled-adder-based constructions. Controlled $n$-qubit add-subtract circuits, which perform an addition when the control qubit is one and a subtraction when it is zero, require only $n-1$ Toffoli gates, instead of the $2n-1$ needed for controlled adders. Despite the existence of  multiplication circuits with better asymptotic scaling, schoolbook multiplication yields the lowest Toffoli counts for small register sizes, making it advantageous in practical applications. For example, the presented approach reduces the Toffoli count by up to around 30\% in circuits for breaking 256-bit elliptic curve keys compared to circuits with standard schoolbook multipliers.

\end{abstract}

In the context of fault-tolerant quantum computation~\cite{Campbell2016,TerhalRMP}, the cost of a quantum circuit is commonly quantified in terms of the Toffoli count, i.e., the number of Toffoli gates in the circuit assuming that the remaining gates are Clifford gates. We focus on quantum circuits that implement the unitary $U$ for quantum multiplication in superposition where both factors are stored in $n$-qubit quantum registers, i.e., $U|x\rangle|y\rangle|0\rangle = |x\rangle|y\rangle|xy\rangle$. We also consider variations of this operation in which a modular reduction is performed on the result. Our results are summarized in Tab.~\ref{tab:results}.

\begin{figure*}[t]
\centering
\includegraphics[width=\linewidth]{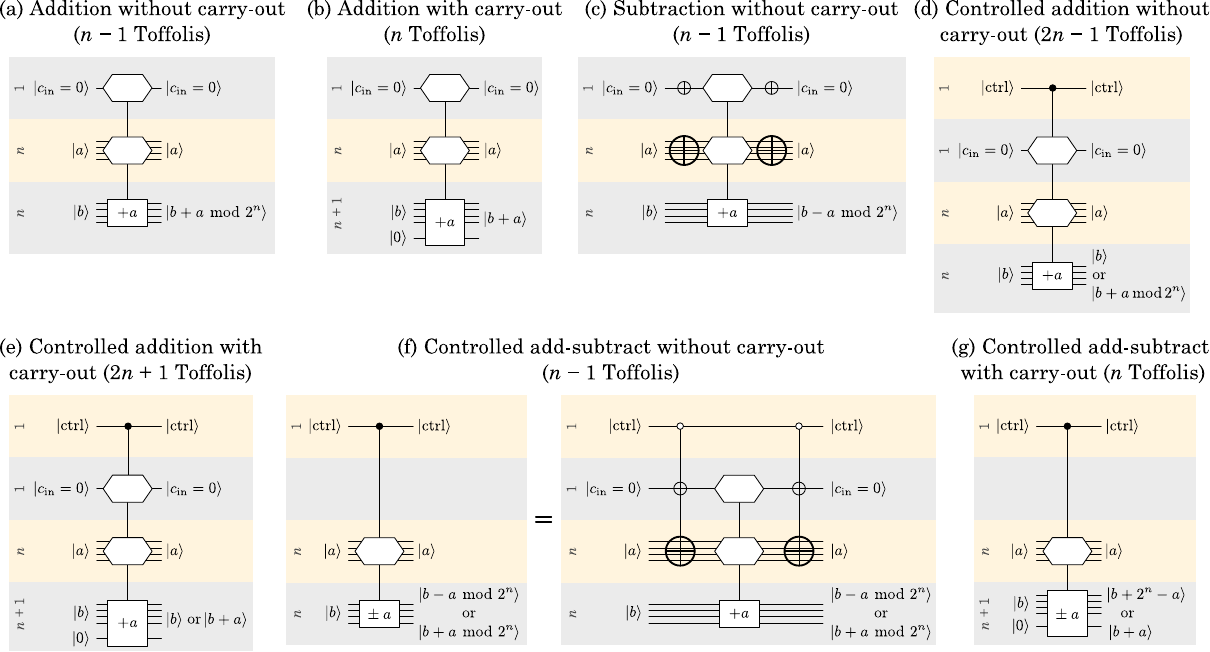}
\caption{Toffoli counts of quantum circuit for addition with (a) and without (b) carry-out, subtraction (c), controlled addition with (d) and without (e) carry-out, and controlled add-subtract with (f) and without (g) carry-out. In these circuits, $|\text{ctrl}\rangle$ refers to control qubits and $|c_{\rm in}\rangle$ refers to input carry qubits.}
\label{fig:addsubtract}
\end{figure*}

\textbf{Controlled add-subtract.} Using Gidney adders~\cite{Gidney2018}, a quantum ripple-carry addition can be performed with $n$~Toffoli gates if the final carry qubit is kept, or with $n-1$ Toffoli gates if it is discarded, which is referred to as addition with and without carry-out in Fig.~\ref{fig:addsubtract}a/b (see also Fig.~\ref{fig:gidneyadder}). Subtraction can be performed using the same circuit by flipping all bits of the subtrahend (Fig.~\ref{fig:addsubtract}c). Similarly, a controlled addition with and without carry-out can be performed with $2n+1$ and $2n-1$ Toffoli gates, respectively (Fig.~\ref{fig:addsubtract}d/e). A controlled adder performs an addition if the control qubit is in the $|1\rangle$ state, and performs no operation if it is in the $|0\rangle$ state. We consider a similar circuit that we refer to as a \textit{controlled add-subtract}. It performs an addition if the control qubit is in the $|1\rangle$ state, and a subtraction if it is in the $|0\rangle$ state. A controlled add-subtract can be implemented with the same Toffoli count as an uncontrolled adder. As shown in Fig.~\ref{fig:addsubtract}f, a controlled add-subtract merely requires two multi-target controlled-NOT gates before and after a standard adder. Therefore, a controlled add-subtract with or without carry-out has a Toffoli count of $n$ or $n-1$ (Fig.~\ref{fig:addsubtract}f/g). If we interpret all numbers in a controlled add-subtract with carry-out as positive integers, then the circuit outputs $b+a$ if the control qubit is $|1\rangle$ and $b + 2^n - a$ if it is $|0\rangle$, as shown in Fig.~\ref{fig:addsubtract}g. Controlled add-subtracts have previously been described in the literature in the context of circuits for square roots~\cite{Munoz2017} and controlled rotations~\cite{Sanders2020}.

\renewcommand{\arraystretch}{1.5}

\begin{table}[t]
\centering
\scalebox{0.92}{
\hskip-0.5cm
\begin{tabularx}{\columnwidth}{ll}
Mult. circuit & Toffoli count \\
\hline
Schoolbook & $n^2 +4n+3$ (compared to $2n^2+n$) \\
Modular mod $2^n$ & $0.5n^2 + 1.5n$ (compared to $n^2$) \\
Modular mod $p$ & $n^2 + 6n + \frac{n}{w}(2^w + 3 \cdot 2^{w/2} + 3n  - 3)$  \\
\multicolumn{2}{r}{(compared to $2n^2 + 4n + \frac{n}{w}(2^w + 3 \cdot 2^{w/2} + n - 1)$)} \\ 
\end{tabularx}}
\caption{Toffoli counts of three different $n$-qubit schoolbook multiplication circuits in this paper: a standard schoolbook multiplication with a $2n$-qubit result register, a modular multiplication circuit in which only $n$ bits of the result are kept, and a modular multiplication circuit in which a modular reduction mod $p$ is performed on the result, where $p$ is a prime number. For the latter, the Toffoli count depends on an integer $w$, which is a tunable \textit{window size}. The optimal value is approximately $w \approx \log_2\left(\frac{n}{\log_2 n}\right) + 2$.}
\label{tab:results}
\end{table}

\begin{figure*}[t!]
\centering
\includegraphics[width=0.9\linewidth]{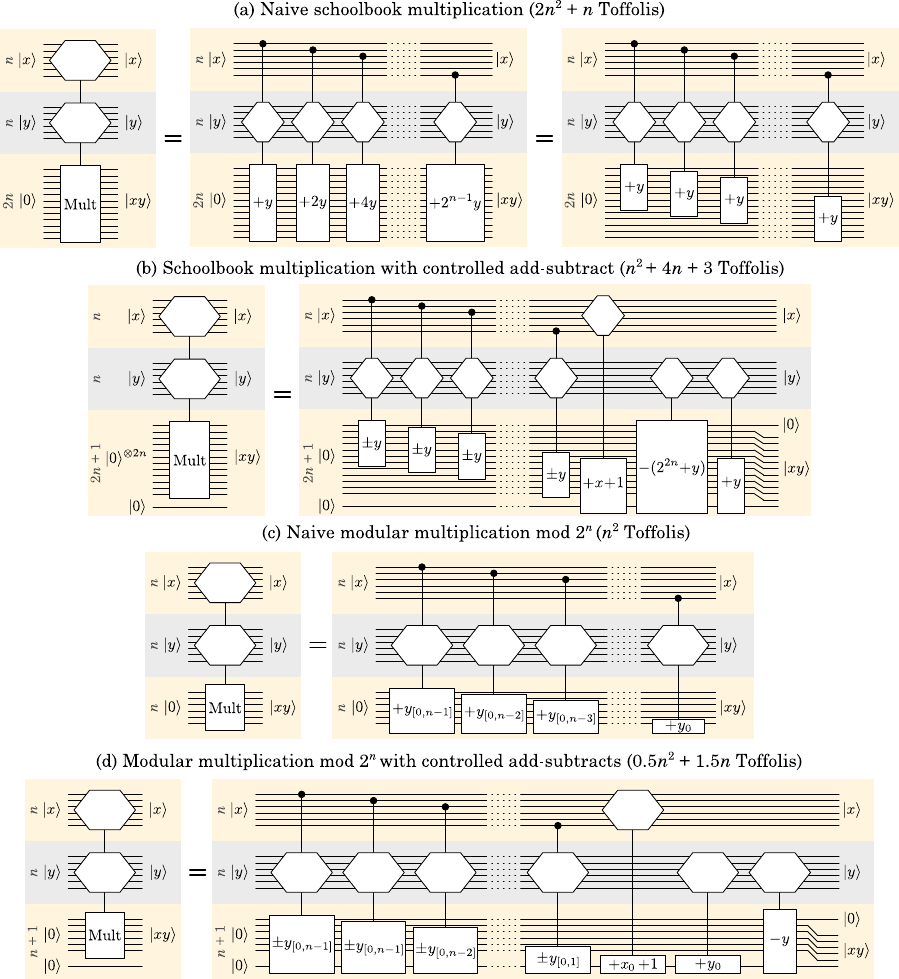}
\caption{Schoolbook multiplication circuits based on controlled adders (a) and controlled add-subtracts (b), and modular multiplication circuits mod $2^n$ based on controlled adders (c) and controlled add-subtracts (d). Here, $y_{[a,b]}$ denotes the bit string formed by extracting the bits of $y$ between the $a$'th and $b$'th least significant bits.}
\label{fig:schoolbook}
\end{figure*}

\textbf{Schoolbook multiplication.} We first consider a multiplication operation that acts on two $n$-qubit registers storing two integers $|x\rangle$ and $|y\rangle$, and computes $|xy\rangle$ in an additional $2n$-qubit result register. Such a schoolbook multiplication circuit can be implemented using the circuit shown in Fig.~\ref{fig:schoolbook}a. The circuit straightforwardly computes the product $xy = \sum\limits_{k=0}^{n-1}x_k \cdot 2^ky$. Each addition of $x_k2^ky$ can be performed with a controlled addition controlled on $x_k$ (the $k$-th bit of $x$ where $x_0$ is the least significant bit) adding $y$ shifted by $k$ bits. Therefore, the circuit consists of $n$ controlled $n$-qubit additions with carry-out, in which case it has a Toffoli count of $2n^2 + n$.

Now consider the circuit in Fig.~\ref{fig:schoolbook}b. If we replace each controlled addition with a controlled add-subtract, each step will be adding $(1-x_k)2^{n+k} + (2x_k-1) \cdot 2^k \cdot y$. In other words, for $x_k=1$, the operation is adding $2^ky$ (as in a standard controlled adder), but for $x_k = 0$, it is adding $2^{n+k} - 2^k y$. After $n$ controlled add-subtract operations, the intermediate result is

$$ \sum\limits_{k=0}^{n-1} (1-x_k)2^{n+k} + (2x_k-1)\cdot 2^k y$$
$$ = 2xy +2^{2n} - 2^{n}(x+1+y)+y \, .$$

Some corrections need to be performed to obtain the intended result $xy$. We first add $2^n(x+1)$ using an $n$-qubit adder with the input carry set to 1. Then we perform a $2n$-qubit subtraction of $2^{2n}+y$, and finally an $n$-qubit addition of $2^ny$. We divide the resulting number by two by simply relabeling the qubits. This results in a schoolbook multiplication circuit with only $n^2 + 4n + 3$ Toffoli gates.

\textbf{Modular multiplication mod $2^n$.} The naive schoolbook multiplication circuit consists of $n$ controlled $n$-bit adders with carry-out. A naive modular multiplication circuit mod $2^n$ can be obtained by replacing the $j$-th controlled adder with a controlled ($n+1-j)$-bit  adder without carry-out, as shown in Fig.~\ref{fig:schoolbook}c. This results in a Toffoli count of $n^2$.

In the circuit in Fig.~\ref{fig:schoolbook}d, we use controlled add-subtracts with carry-out instead of controlled adders without carry-out. The intermediate result is a $2n+1$-bit number corresponding to the intermediate result of the schoolbook multiplication mod $2^{n+1}$, i.e.,

$$2xy +2^{2n} - 2^{n}(x+1+y)+y \mod 2^{n+1} $$
$$ = 2xy -2^n(x+1+y) + y \mod 2^{n+1} \, .$$

The required correction operations are cheaper compared to the previous schoolbook multiplication, as they are performed modulo $2^{n+1}$. This turns the addition of ${2^n(x+1)}$ into an addition of $x_0 + 1$, where $x_0$ is the least significant bit of $x$. As this corresponds to a controlled-NOT gate conditioned on $x_0$ followed by a NOT gate, it does not contribute to the Toffoli count. Similarly, the addition of $2^n y$ turns into an addition of $y_0$. The subtraction of $2^{2n}+y$ turns into a subtraction of $y$, which contributes $n$ Toffoli gates to the total Toffoli count of $0.5n^2 + 1.5n$.

\begin{figure*}[t]
\centering
\includegraphics[width=\linewidth]{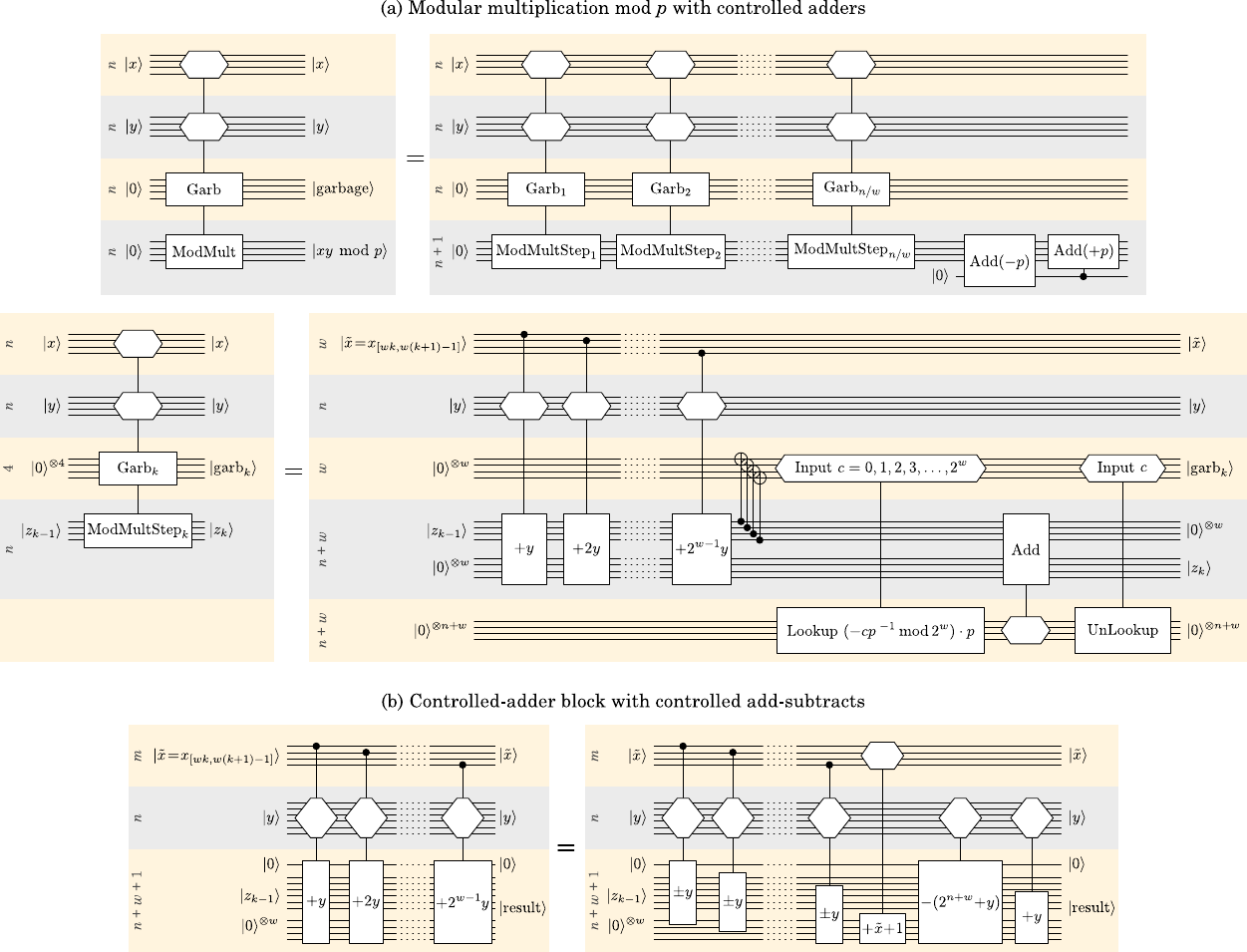}
\caption{Modular multiplication circuits mod $p$ based on controlled adders~\cite{Haener2020, Gouzien2023} (a) and controlled add-subtracts (b).}
\label{fig:modmult2}
\end{figure*}

\textbf{Modular multiplication mod $p$.} An out-of-place modular multiplication construction is described in Refs.~\cite{Haener2020, Gouzien2023} for the case that the modulus is a prime number $p$. It assumes that the factors are stored in Montgomery representation~\cite{Montgomery1985}. The circuit is shown in Fig.~\ref{fig:modmult2}a. It consists of $n/w$ $\text{ModMultStep}$ operations that each contribute $w$ qubits to a garbage register, followed  by two additions of a constant for modular reduction each contributing $n$ Toffoli gates. If necessary, the garbage register can be uncomputed by copying the multiplication result into a separate register using controlled-NOT gates, and then uncomputing the entire multiplication. Each $\text{ModMultStep}$ consists of $w$ controlled additions with carry-out contributing $2w(n+1)$ Toffoli gates, a $2^w$-item lookup table contributing $2^w$ Toffoli gates~\cite{Babbush2018}, an $n+w$-qubit adder contributing $n+w-1$ Toffoli gates, and an uncomputation of the lookup table contributing $3\sqrt{2^w}$ Toffoli gates~\cite{Gidney2019b}. This results in a total Toffoli count of $2n^2 + 4n + \frac{n}{w}\left(2^w + 3\cdot 2^{w/2} + n -1 \right)$. The choice of $w$ that minimizes the Toffoli count is approximately $w \approx \log_2\left(\frac{n}{\log_2 n}\right) + 2$.

The first half of each $\text{ModMultStep}$ consists of $w$ controlled additions that are controlled on the $w$ bits of $x$ starting with the $wk$-th least significant bit. We refer to this number as $\tilde{x} = x_{[wk, w(k+1)-1]}$. The input to the circuit is an $n$-qubit number $z_{k-1}$ and the circuit outputs an $n+w$-qubit number

$$ \mathrm{result} = z_{k-1} + \sum\limits_{k=0}^{w-1} \tilde{x}_k \cdot 2^ky \, .$$

We construct a version with controlled add-subtracts in Fig.~\ref{fig:modmult2}b. Here, we first double the input number $z_{k-1}$ by adding a 0 as the least significant bit. We treat $y$ as an $n+1$-qubit number (with the most significant bit set to 0) and perform controlled add-subtracts to obtain an intermediate result

$$ 2z_{k-1} + \sum\limits_{k=0}^{w-1}(1-\tilde{x}_k)2^{n+k} + (2\tilde{x}_k-1)\cdot 2^ky $$
$$ = 2 \cdot \mathrm{result} - 2^n(\tilde{x}+1)+2^{n+w}+y-2^wy \, .$$

The controlled-add-subtracts contribute $w(n+1)$ Toffoli gates. We apply the required corrections by performing a $w$-qubit addition to add $2^n(\tilde{x}+1)$ with a Toffoli cost of $w-1$, an $n+w+1$-qubit subtraction to subtract $2^{n+w}+y$ with a Toffoli cost of $n+w$, and finally an $n$-qubit addition to add $2^wy$ with a Toffoli cost of $n-1$. We obtain the final result after performing a division by two by relabeling qubits. The total Toffoli count of the multiplication is $n^2 + 6n + \frac{n}{w}\left(2^w + 3 \cdot 2^{w/2} + 3n - 3\right)$.

\textbf{Conclusion.} Controlled add-subtracts can be used to construct multiplication circuits with half the asymptotic Toffoli count compared to constructions based on standard controlled adders. The add-subtract-based constructions outperform the controlled-adder-based ones for all $n \geq 4$. However, the Toffoli count reduction only starts exceeding 25\% at $n = 8$ for schoolbook multiplication, at $n = 6 $ for modular multiplication mod $2^n$, and at $n = 65$ for modular multiplication mod $p$. Still, even the latter can be useful in practical applications. For example, several implementations of Shor's algorithm for breaking elliptic curve keys~\cite{Shor1994,Shor1997, Proos2003, Roetteler2017, Haener2020, Gouzien2023, Litinski2023} primarily rely on modular multiplication. The implementation in Ref.~\cite{Litinski2023} is a family of circuits that can be used to break 256-bit keys using $(44 + 65/k) \cdot 10^6$ Toffoli gates with around $3000k$ logical qubits of memory, where $k$ is an integer. Replacing the controlled-adder-based multiplication circuits with controlled-add-subtract-based circuits reduces the Toffoli count to $(30 + 73/k) \cdot 10^6$ Toffoli gates per key. Asymptotically, this is a reduction of over 30\% from 44 million to 30 million Toffoli gates per key. However, for moderate values of $k$, this reduction will be lower, e.g., a reduction of 20\% for $k=4$, going from 60 million to 48 million Toffoli gates per key.

Multiplication circuits are also useful outside of cryptography applications. For instance, in some variations of quantum simulation algorithms, they may be used to compute potential and kinetic energies. While there exist multiplication circuits with better asymptotic scaling~\cite{Parent2017, Dutta2018, Gidney2019c, Kahanamoku2024}, schoolbook multiplication is typically still cheaper for practically relevant problem sizes. This paper is also meant to highlight the usefulness of the controlled add-subtract operation. It has previously been used to reduce the cost of square roots~\cite{Munoz2017} and controlled rotations~\cite{Sanders2020}, so perhaps it can be used to reduce the cost of other subroutines that rely on controlled adders.

\section*{Acknowledgments}

I would like to thank Jessica Lemieux and Mark Steudtner for discussions that inspired this construction.

\appendix

\section{Additional figures}

\begin{figure*}[t]
\centering
\includegraphics[width=\linewidth]{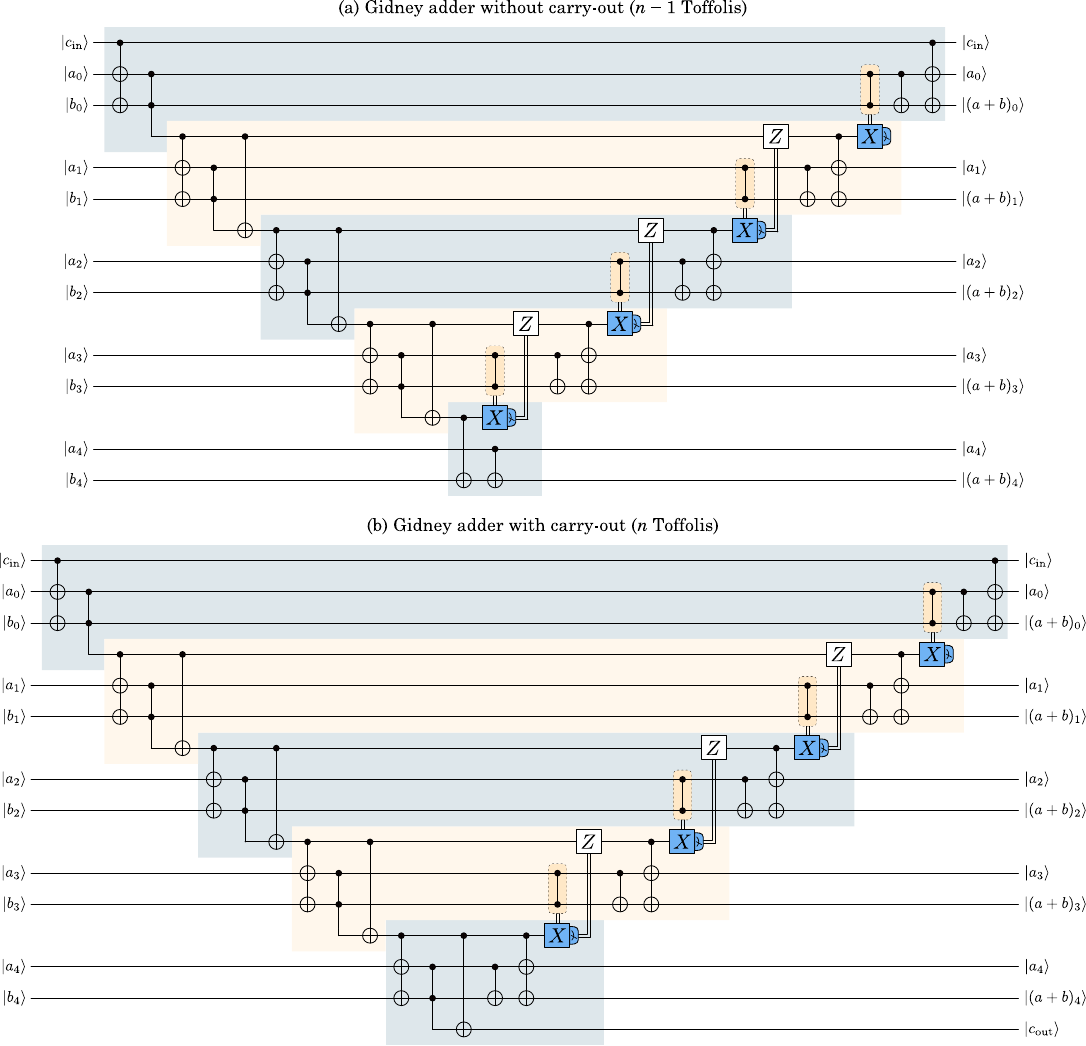}
\caption{Examples of Gidney adders~\cite{Gidney2018} with and without carry-out for $n=5$.}
\label{fig:gidneyadder}
\end{figure*}

Figure \ref{fig:gidneyadder} shows a Gidney adder~\cite{Gidney2018} with and without carry-out.

\end{document}